\documentclass[amsmath,amssymb,aps,pre,twocolumn]{revtex4-2}
\usepackage[utf8]{inputenc}
\usepackage[T1]{fontenc}
\usepackage{amsmath}
\usepackage{graphicx}
\usepackage{booktabs}
\usepackage{adjustbox}
\usepackage{hyperref}
\usepackage{cancel}
\usepackage{bm}
\usepackage{mathtools}
\usepackage{svg}
\hypersetup{colorlinks,
	citecolor=blue
}
\usepackage{natbib}
\usepackage{multibib}
\usepackage{bbm}
\usepackage{hhline}
\usepackage{float}
\graphicspath{ {Images/} }
\usepackage{array}
\usepackage{array,multirow}
\newcites{sm}{\LaTeX-Literature}
\usepackage{bbm}
\graphicspath{ {Images/} }
\usepackage{array}
\newcolumntype{P}[1]{>{\centering\arraybackslash}p{#1}}

\usepackage{times}
\usepackage{setspace}
\usepackage{txfonts}
\usepackage{pxfonts}
\usepackage{todonotes}

\definecolor{BramGreen}{rgb}{0.0, 0.5, 0.0}

\begin{document}

\title{Nonlinear random walks optimize the trade-off between cost and prevention in epidemics lockdown measures : the \textit{ESIR} model}
\author{Bram A. Siebert} \affiliation{MACSI, Department of Mathematics and Statistics, University of Limerick, Limerick V94 T9PX, Ireland} 
\author{James P. Gleeson} \affiliation{MACSI, Department of Mathematics and Statistics, University of Limerick, Limerick V94 T9PX, Ireland}
\author{Malbor Asllani} \affiliation{School of Mathematics and Statistics, University College Dublin, Belfield, Dublin 4, Ireland}

\date{\today}

\begin{abstract}
\noindent
Contagious diseases can spread quickly in human populations, either through airborne transmission or if some other spreading vectors are abundantly accessible. 
They can be particularly devastating if the impact on individuals' health has severe consequences on the number of hospitalizations or even deaths. 
Common countermeasures to contain the epidemic spread include introducing restrictions on human interactions or their mobility in general which are often associated with an economic and social cost. 
In this paper, we present a targeted model of optimal social distancing on metapopulation networks, named $ESIR$ model, which can effectively reduce the disease spreading and at the same time minimize the impact on human mobility and related costs. 
The proposed model is grounded in a nonlinear random walk process that considers the finite carrying capacity of the network's metanodes, the physical patches where individuals interact within mobility networks. This later constrain is modeled as a \emph{slack compartment} $E$ for the classic $SIR$ model and quantifies the density of vacant spaces to accommodate the diffusing individuals.
Formulating the problem as a multi-objective optimization problem shows that when the walkers avoid crowded nodes, the system can rapidly approach Pareto optimality, thus reducing the spreading considerably while  minimzing the impact on human mobility as also validated in empirical transport networks. 
These results envisage \emph{ad hoc} mobility protocols that can potentially enhance policy making for pandemic control.   
\end{abstract}

\maketitle

\section{Introduction}

\noindent
Epidemics have affected humankind many times during history with countless casualties and enormous economic and social impact. 
For example, the notorious Black Death (1331--1353) caused 75--200 million deaths worldwide and led to the loss of life of approximately one third of Europe's population \cite{benedictow_black_2004}. At the time of writing, the COVID-19 pandemic has drastically influenced our lives and affected our well-being in many aspects, creating a real challenge in finding a compromise between curbing the spread of the virus and allowing an adequate level of normality in human activity. 
Human behavior is one the driving factors of epidemic spreading and thus has long been the focus of scientists, governments, and public health officials \cite{Colizza2006, Balcan2009}.
Many attempts have been made to model measures taken by the public, such as social distancing \cite{Funk2010, Perra2011}.
The recent Coronavirus pandemic has reawakened the necessity for a thorough understanding of how to control the spread of a highly contagious disease, with many novel studies emerging recently \cite{Kraemer2020, Vespignani2020, Fang2020}.
Many governments have taken measures to slow down the spread of the COVID-19 virus by restricting, and thus reducing, human mobility \cite{Aleta2020_1, Aleta2020_2, Liu2020}.
While this has successfully contained the spreading, it has come at an enormous economic and social cost, for example,  the complete closure of schools, non-essential retail, manufacturing plants, etc.
Furthermore, stopping non-essential medical procedures will unavoidably cause long-term health issues not related to COVID-19 \cite{Bonaccorsi2020}.

The spreading of diseases has been traditionally described using continuous differential equations such as reaction-diffusion models in spatially extended systems \cite{Petard1927, volpert_reactiondiffusion_2009}.
These models assume that the human population diffuses homogeneously in some continuous domain, and contact occurs once two or more individuals are sufficiently near to each other \cite{Britton2020}. 
This perspective, however, is far from being realistic, in particular because people moving from one spatial patch to another follow well-defined paths that a continuous spatial domain cannot capture. 
On the other hand, the interactions people have once they are in the vicinity can hardly be considered homogeneous or well-mixed.  
Inspired by that, in recent years, new network-based reaction-diffusion models have been developed \cite{Newman2010, Barthelemy2011, Colizza2007, Belik2011, Chinazzi2020} and significant efforts in understanding the structure of human populations and their movement have been made with critical outcomes in understanding how an epidemic spreads \cite{Funk2010}.
From this novel perspective, network nodes represent the different spatial patches where individuals can interact, and  network edges denote the paths through which they can move. 
Modelling and detecting individuals' movements has been a major challenge for researchers in human dynamics \cite{gonzalez_understanding_2008, song_limits_2010}. 
Although it might resemble an oversimplification, random walks are an excellent first approximation to the diffusion of individuals of a sufficiently large population \cite{song_modelling_2010, barbosa_human_2018}.  

\begin{figure}[h!]
    \centering
    \includegraphics[width=\columnwidth]{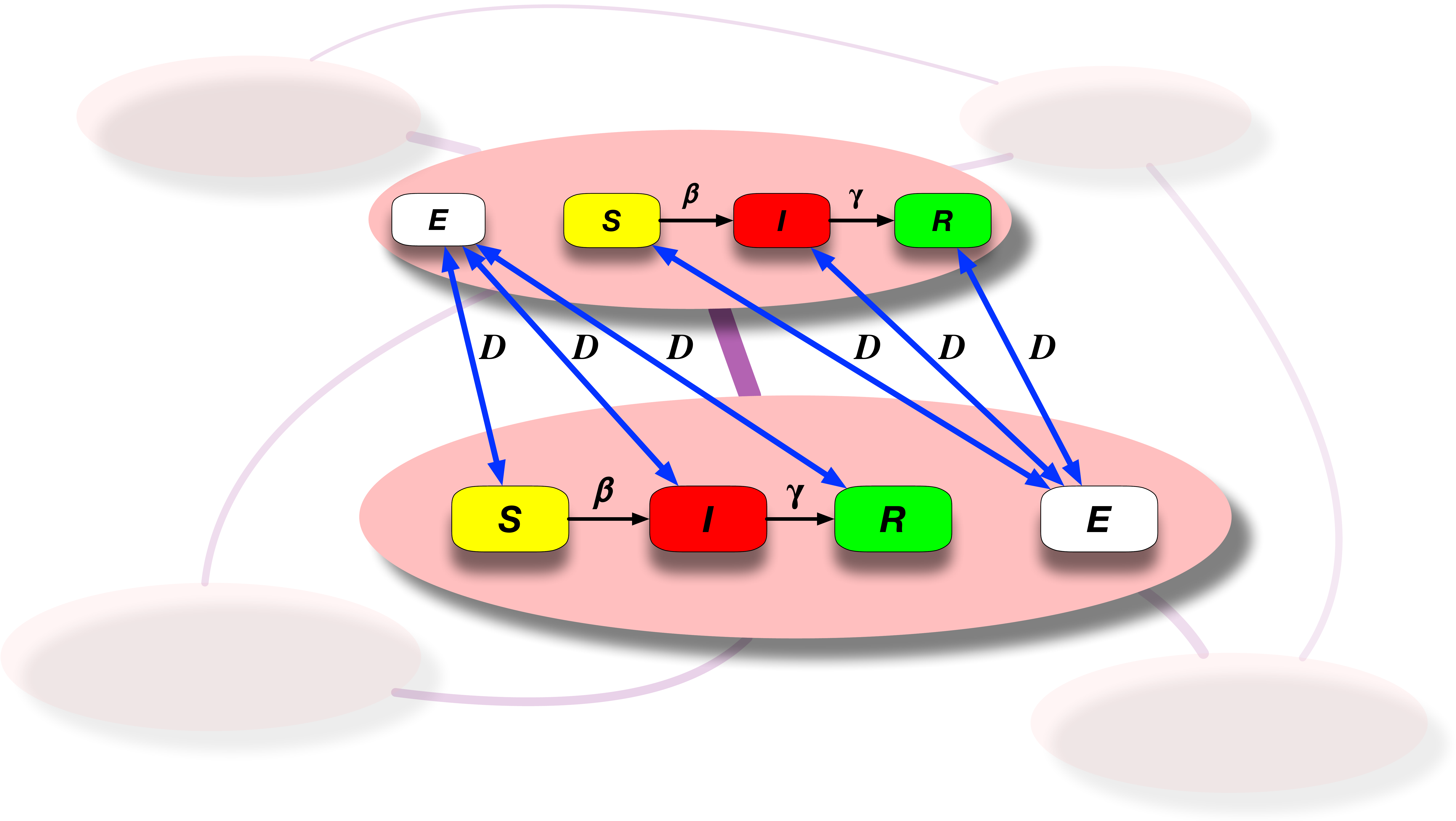}
    \caption{A schematic representation of the $ESIR$ model across two nodes of the network. The model consists of a classic $SIR$ model in each of the nodes (pink ovals), complemented with the Empty compartment $E$. The dynamics between the different compartments are $S_i\xrightarrow{\beta}I_i\xrightarrow{\gamma}R_i$ for each node $i$, indicated by the black solid arrows and where $\beta$ and $\gamma$ represent respectively the contagion rate and the recovery rate. In addition, we have also considered (blue arrows) the internode diffusion relations $S_i\xleftrightarrow{D}E_j$, $S_j\xleftrightarrow{D}E_i$, $I_i\xleftrightarrow{D}E_j$, $I_j\xleftrightarrow{D}E_i$, and $R_i\xleftrightarrow{D}E_j$, $R_j\xleftrightarrow{D}E_i$ where $D$ represent the diffusion constant and $i$ and $j$  are the connected nodes. The double arrows in the these transitions mean that when an individual moves towards an adjacent node it occupies a vacant space and in turn creates an empty space in the origin node.}
    \label{fig:cartoon}
\end{figure}

Based on this synopsis, in this work, we present a reaction-diffusion model that aims to slow down the spread of infection through a targeted reduction of mobility as opposed to a global restriction on it. 
At the same time, our goal is to reduce the adverse effects that a partial or complete lockdown may cause.  
At odds with the traditional random walk diffusion, where individuals diffuse through the network in an ``unselective'' fashion, in our case, we have modified the rules such that individual walkers choose which node to hop into next.
Such biased random walk diffusion is based on recent developments that take into account that each node (i.e., spatial patch) has a finite-size carrying capacity \cite{Asllani2018, Carletti2020}.
This constraint adds a nonlinear term to the simple (unbiased) random walk process, yielding a different steady-state configuration. 
In other words, such a new ``microscopic'' movement rule forces the diffusing individuals to avoid nodes where there is already a high density of individuals, preferring less crowded ones instead.
Such a phenomenon when correlated with the local interactions at the node level restricts the spreading of the epidemic without necessarily increasing the economic/social cost, usually an unavoidable consequence of mobility reduction. 
In fact, the finite carrying capacity of the nodes can be considered an extra compartment that can complement the already existing epidemiological models.
We will denote this compartment as $E$, which stands for the \emph{Empty} compartment. 
Therefore, the classic Susceptible-Infected-Recovered ($SIR$) model that we will consider throughout this paper now becomes an $ESIR$ compartment model (see Fig. \ref{fig:cartoon} for a schematic presentation. Let us also notice that such a new slack compartment is artificial and does not represent any particular type of individual but rather the density of vacant space that the individuals can occupy.
Following this formalism, the conservation property for the $ESIR$ model is written as $E+S+I+R=1$ where the symbols $S$, $I$, $R$, and $E$ will represent the densities of the respective species of individuals and the empty space.
Grounded in the Pareto efficiency formalism and using the Entropy Rate \cite{cover_elements_2006} to quantify the average process performance, in this work, we show that it is possible to reduce the infection in the population by redistributing the individuals differently among the nodes while still keeping the same average mobility.     
Therefore, our model indicates that more specific and targeted policies can be more effective in controlling the spreading of diseases and simultaneously diminish the adverse effects of reduced mobility.

\onecolumngrid

\twocolumngrid

\vspace{10cm}

\section{Nonlinear random walks with finite-size nodes}

A metapopulation network consists of nodes representing spatial patches where groups of individuals congregate \cite{Calvetti2020}. 
The nodes are connected through preferential paths known as edges through which the individuals diffuse. 
In order to model the diffusion process in a metapopulation network it is crucial to consider the system's overall behaviour without excluding any of the walkers.   
To this aim, the master equation is responsible for the definition of the stochastic dynamics that governs the global distribution of the individuals for each node of the network at any instant of time \cite{gardiner_handbook_2004, kampen_stochastic_2007}
\begin{equation}
    \frac{d}{dt}P(\textbf{m},t) = \sum_{\textbf{m}'}\big[T(\textbf{m}|\textbf{m}')P(\textbf{m'},t) - T(\textbf{m}'|\textbf{m})P(\textbf{m},t)\big].
\end{equation}
Here $P(\textbf{m},t)$ denotes the probability that the system is in state $\textbf{m}$ at time $t$, $T(\textbf{m}|\textbf{m}')$ is the transition probability from state $\textbf{m}'$ to $\textbf{m}$, and the sum is restricted to states which $\textbf{m}'$ can transition from \cite{Asllani2018}.
Since we will initially limit our discussion to the diffusion process only, the individual-based dynamics considers the hopping of walkers from one node to another.
Thus with a little abuse in notation the transition probabilities are
\begin{equation}\label{eq:Master}
    T(m_i-1,m_j+1 | m_i,m_j) = \frac{A_{ij}}{k_i}f\left(\frac{m_i}{M}\right)g\left(\frac{m_j}{M}\right),
\end{equation}
where $m_i$ are elements of the vector $\mathbf{m}$, $M$ is total number of sites available per node (that for simplicity we consider to be the same for all nodes), $A_{ij} = 1$ if there is a link between nodes $i$ and $j$, $k_i = \sum_j A_{ij}$, is the degree of node $i$, and $f(\cdot)$ and $g(\cdot)$ are two nonlinear functions representing the will of each individual to leave node \emph{i}, and  the will to settle in node \emph{j} respectively. 
To keep the discussion simple, throughout this paper, we will define these functions depending on the concentration ${m_i/M}$ as $f(m_i/M)=m_i/M\,$, and $g(m_i/M)=(1-m_i/M)^{\sigma_i}$,
where the parameter $\sigma_i$ represents the amount of congestion at node $i$, that we will generally refer to as the\emph{crowding} parameter.
Recall that the term $1-m_i/M$ represents the density of the empty space in node $i$.
It is easy to see that in the limit, ${\sigma_i\rightarrow 0,\,\forall i}$, the random walk with crowding collapses to the simple random walk case. 
Before we proceed further, let us emphasise that the crowding phenomenon in terms of molecular or cellular level occurs for ${\sigma_i=1,\,\forall i}$ \cite{liggett_stochastic_1999, neri_totally_2011}. 
In fact, from the physical point of view, particles are thought to displace proportionally to the vacant space of the hosting nodes, an example of interacting particles systems known as asymmetric simple exclusion process (ASEP) \cite{liggett_stochastic_1999}. 
However, when we extend the idea of the biased random walkers to the case of humans (the same consideration can also hold for animals), their perception of \emph{attractiveness} $g(\rho_j)$ of the hosting node $j$, can be different from the amount of the available space in such node, and the crowding parameter $\sigma_j$ aims to quantify such relation.
We believe that this formulation better reflects the properties of human mobility where the gathering in common areas (e.g., schools, shops, offices, etc.) depends on the individual's perception of the attractiveness of vacant space.

Although equation \ref{eq:Master} is exact in describing the dynamics, it is, unfortunately, impractical to deal with. 
Thus, to describe the time-evolution of the node density $\rho_i(t)$, the implementation of a mean-field approach is required. 
Based on that, we first average over the different configurations ${\langle m_i \rangle = \sum_\textbf{m} m_i P(\textbf{m},t)}$ and by then taking the limit we obtain the node densities ${\rho_i=\lim\limits_{M\rightarrow \infty}\langle m_i \rangle/M}$. 
In conclusion, we obtain the mean-field differential equations
\begin{equation}\label{eq:meanFieldDiffusion}
    \frac{d\rho_i}{dt}= \sum_{j=1}^{\Omega} \Delta_{ij} \left[ f(\rho_j)g(\rho_i) - \frac{k_j}{k_i}f(\rho_i)g(\rho_j) \right], \quad \forall i
\end{equation}
where $\Omega$ is the number of nodes in the network, ${\Delta_{ij} = A_{ij}/k_j - \delta_{ij}}$ is the Linear Random Walk (LRW) Laplacian, and time is rescaled so that ${t/M \mapsto t}$. 
Notice that in the limit of large $M$ we can neglect any correlation ${\langle f(\cdot)g(\cdot) \rangle = f(\langle \cdot \rangle) g(\langle \cdot \rangle)}$ following the Van Kampen ansatz \cite{gardiner_handbook_2004, kampen_stochastic_2007}. For further details on the derivation and related generalisations the interested reader can refer to \cite{Carletti2020}. To make eq. \eqref{eq:meanFieldDiffusion} more compact and to reflect our choice for the functions $f(\cdot)$ and $g(\cdot,\cdot)$, we introduce the nonlinear diffusion operator 
\begin{equation*}
    L_{ij}(\rho) = \Delta_{ij} \left[\rho_j (1-\rho_i)^{\sigma_i} - \frac{k_j}{k_i} \rho_i (1-\rho_j)^{\sigma_j}\right], \label{eq:Laplacian} 
\end{equation*}
referring to it as the Nonlinear Random Walk (NLRW) operator, a notation that we will use for the rest of this paper. 

From the mean-field perspective, although we are modeling human populations and will refer to members as individuals, it is essential to note that the node populations (densities) are not discrete but rather continuous. 
The continuous approximation allows us to easily model individuals' movement through the network and is computationally more convenient, thanks to the deterministic (PDE) formulation. 
Also, from eq. \eqref{eq:meanFieldDiffusion}, one can immediately notice that at variance with the simple random walk diffusion, the diffusion in our model depends on the state ${g(\cdot)}$ of the hosting node. 
This function describes the available space in the destination node, which can be filled by individuals coming from the origin node.

\section{Epidemic modelling in crowded networks}

We will now augment the above formalism to consider also the local contagion dynamics at the node level. 
To this aim, we will refer to the celebrated Susceptible-Infected-Recovered ($SIR$) model \cite{murray_mathematical_2002, Kiss2017}, a 3-compartment model, where the (healthy) susceptible individuals, $S$, can potentially become infected with probability $\lambda$ if they come in contact with infected individuals $I$.  
On the other side, infected individuals become recovered, $R$, at a rate $\gamma$ and participate in the diffusion dynamics as immune to the infection.   
Putting together the nonlinear random walk diffusion of the individuals between adjacent nodes, with the contagion dynamics occurring within nodes, we have the new mean-field set of equations, 
\begin{eqnarray}
    \frac{dS_i}{dt} &=& -\lambda S_iI_i + D\sum_{j=1}^\Omega L_{ij}(S), \nonumber \\
    \frac{dI_i}{dt} &=& \lambda S_iI_i -\gamma I_i + D\sum_{j=1}^\Omega L_{ij}(I), \\
    \frac{dR_i}{dt} &=& \gamma I_i + D\sum_{j=1}^\Omega L_{ij}(R), \nonumber
    \label{eq:SIR}
\end{eqnarray}
where $S_i$, $I_i$, and $R_i$ represent the density of susceptible, infected, and recovered individuals respectively in node $i$, $\lambda$ is the infection rate, and $\gamma$ is the recovery rate. The rates are defined under the standard assumption of well-mixed populations within each node.
Finally, $D$ denotes the diffusion constant, which for simplicity is considered to be the same for susceptible, infected and recovered individuals, as well as equal for all nodes. 
Since the (nonlinear) diffusion is a conservative process \cite{Asllani2018, Carletti2020}, the total number of individuals, remains constant over the network, similar to many other epidemic models \cite{Kiss2017}. 
Nevertheless, it is essential to emphasize that in our model, the densities of the individuals do not sum to unity. In fact, in the Eqs. \eqref{eq:SIR} above, we have intrinsically introduced through the diffusion operator $L$ another compartment $E$ to quantify the density of the vacant space available in each node. Thus in our case the conservation property for each node reads like $S_i+I_i+R_i+E_i=1$.
We name this  model the $ESIR$ \emph{compartment model}. However, we note that it is possible to use other compartment models for more complex spreading mechanisms by simply augmenting them with the $E$ compartment as above.
However, since our focus is on the role that the spatial interactions have on spreading the infection rather than the contagion dynamics, we will constrain our discussion to the $ESIR$ model above. 

\begin{figure}[h!]
\centering
    \includegraphics[width=\columnwidth]{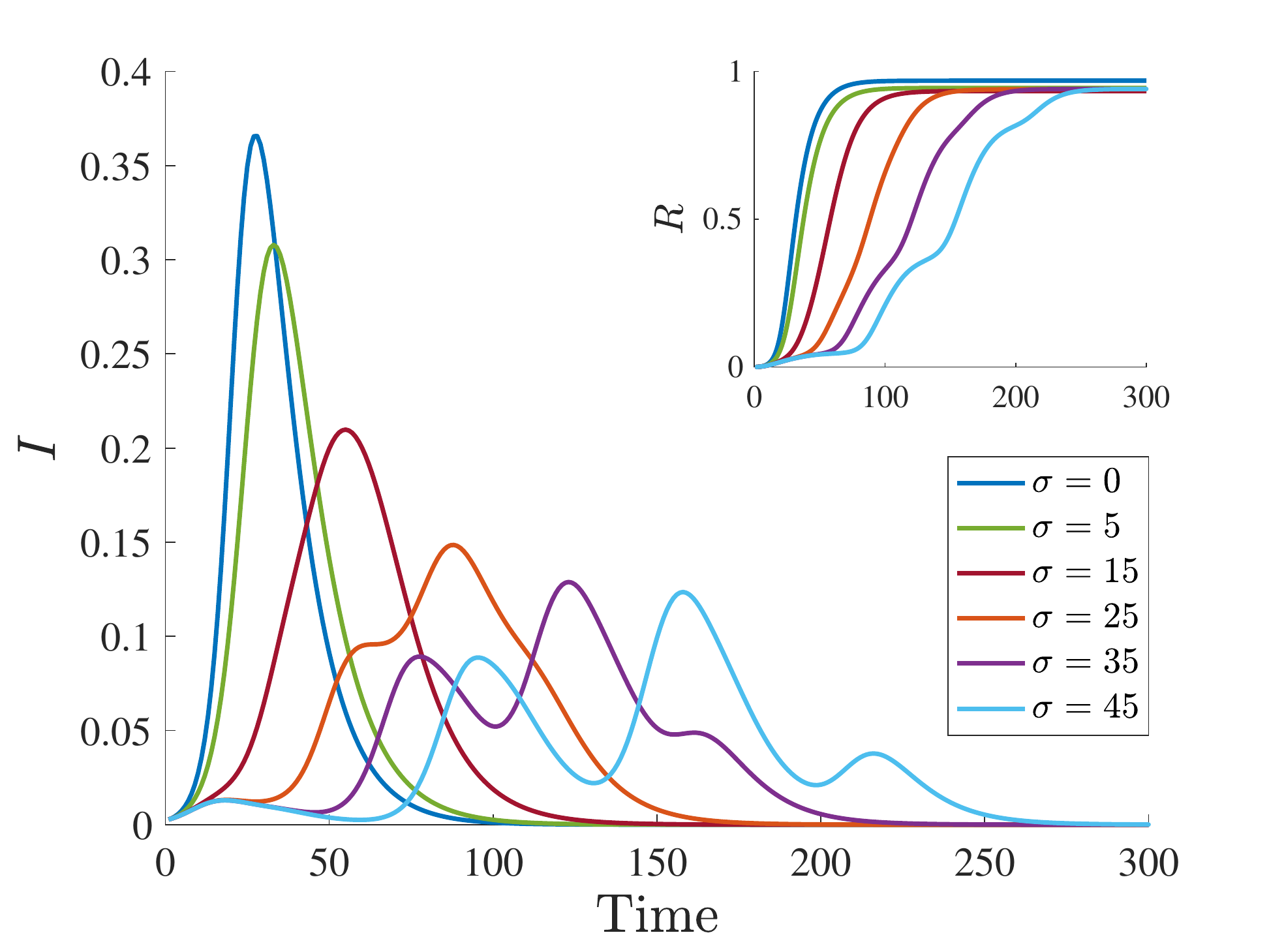}
    \caption{Plots of the average node-level infection $\sum_i I_i/(\Omega\beta)$ and the average node-level recovery $\sum_i R_i/(\Omega\beta)$ versus time, on a latent spatial graph for different values of the uniform crowding parameter $\sigma$. Notice that increasing the crowding parameter $\sigma$ flattens the infection curves and breaks the emerging of the infection peak leading to second and third waves. The latent spatial graph consists of $100$ nodes on a $7.5$ by $7.5$ Euclidean space, $\alpha=0.3$, and the other parameters of the model are  $\lambda = 1$, $\gamma = 0.1$, $\beta = 0.3$, $D = 2$. 
    The infection is seeded at $5$ randomly selected nodes.
    These nodes have $S_i = 0.25$, and $I_i = 0.05$, ensuring $S_i + I_i = \beta = 0.3$.}
    \label{fig:InfectionER}
\end{figure}

To shed light on how the perception of crowding nodes affects the spreading of infection, we will consider a synthetically generated spatial network, known as the \textit{latent spatial} network model \cite{raftery_fast_2012,raftery_comment:_2017}. 
It can be briefly described as an algorithm for generating random geometric graphs; that is, random networks embedded in a Euclidian space where the probability of two given nodes $i, j$ being connected is given by
\begin{equation*}
p(i, j)= 1\big/{\left(1 + e^{-\alpha + d\left(i, j \right)}\right)},
\end{equation*}
where $\alpha$ is a parameter of the model and $d$ the Euclidian distance between the nodes, $d= \big| x(i)-x(j)\big|$, with $x(i)$ representing the coordinates of node $i$. 
Thus, nodes that are closer together in this Euclidean space are more likely to be connected.
The outcome of an infection spreading simulation in the new context of the $ESIR$ model is shown in Fig. \ref{fig:InfectionER} where for simplicity we have fixed the crowding parameter to be equal for every node $\sigma_i=\sigma$. 
It can be observed that increasing the value of $\sigma$, namely the perception of congestion in the hosting nodes, leads to a dramatic decrease in the peak of the infection curves.
This slowing down of the spreading of infection is also associated with multiple waves of infections as well as a general delay in the infection surge.
These features of the infection curves are crucial in addressing hospitalisation capacities.
Such an outcome can intuitively be explained because a significant value of $\sigma$ would slow down the diffusion and decrease mobility, which further isolates the infections and consequently the spreading. Following this logic, we analyse eqs. \eqref{eq:SIR} by linearly expanding for early times of the spreading (see the Appendix for details) for the particular case of regular graphs and $\sigma_i=\sigma, \forall i$. In conclusion, the linearised diffusion operator is equivalent to, for instance for the susceptible individuals
\begin{equation}
D\left(1-S^*\right)^{\sigma} \sum_{j=1}^\Omega\Delta_{ij} S_j
\label{eq:lin_diff}
\end{equation}
as expressed in terms of the LRW Laplacian $\Delta_{ij}$ and where $S^*$ is the initial uniformly distributed fixed point for the susceptible individuals in absence of infected and recovered individual $I^*=R^*$=0. 
Notice that since, in general, the nodes will not be entirely occupied, the term $1-S^*\leq 1$, thus justifying the consistency of the expression above. Equation \eqref{eq:lin_diff} shows that in the absence of the crowding effect $\sigma=0$ (e.g., very diluted conditions $M\rightarrow\infty$), the infection spreads as if being driven by a LRW Laplacian operator. 
However, once we take into consideration the finite capacity of the nodes, with $\sigma>1$, the linear diffusion will slow down. 
\footnote{In principle, it is possible to consider the case $\sigma<1$, which can be interpreted as the scenario when the individuals underestimate the occupancy of the adjacent nodes. However, this case does not significantly affect the overall dynamics, therefore it has not been considered in this study.}
It is important to emphasise that the linear diffusion can only predict the dynamical outcome of the model shortly after the system is initialised, but not for longer time periods. 
Nevertheless, it is reasonable to think that if the diffusion of infection has been limited at early stages, it should decisively influence both the peak of infection and delay the surge of disease in general.
On the other side, it is rational that a drastic decrease in the pace of diffusion of both infected or healthy individuals (the recovered ones are passive in the contagion process) should also strictly reduce the general mobility in the network. 
Next, we will show that it is possible to drastically reduce the overall infection while maintaining high levels of mobility.

\section{Preventing the infection spread while reducing the mobility cost}

Although curbing the peak of infection curve is a highly desired goal of  epidemic control policies, it is often associated with considerable unwanted effects such economic, social and ironically also health costs \cite{pike_economic_2014, fan_pandemic_2018}.
A real challenge from this perspective is to cut down the bulk of the infection and at the same time to keep 
reasonable levels of economic and social activity, which are also necessary for mental well-being. 
Before we show that the $ESIR$ model we propose succeeds in reaching this goal, we first need to introduce a measure which quantifies the level of mobility of active individuals in the network spatial domain. 
To this aim we will refer to an entropic measure borrowed from information theory, known as the  Entropy Rate (ER) \cite{cover_elements_2006}, which has been successfully used to assess the mobility of biased random walkers in complex networks \cite{Gomez-Gardenes2008}. 
For a static Markov chain the ER is defined as $H=-\sum_{ij} \rho_i^* T_{ij}\log T_{ij}$, where $\{\rho_i^*\}$ are the equilibrium states and $\textbf{T}=\{T_{ij}\}$ is the transition matrix evaluated at equilibrium.
In our case this formula is explicitly written as follows (see also \cite{Carletti2020})
\begin{equation} 
H = -D\sum\limits_{ij} \rho_i^*A_{ij}\frac{\rho_i^*\left(1-\rho_j^*\right)^{\sigma_j}}{k_i} \log\left[ A_{ij}\frac{\rho_i^*\left(1-\rho_j^*\right)^{\sigma_j}}{k_i} \right]
\label{eq:entropy}
\end{equation}
where it is noted that the ER depends on the the diffusion and crowding parameters, $D$ and $\sigma_i, \forall i$, respectively, and on the network topology via the adjacency matrix $A=\{A_{ij}\}$. 
The equilibria of the mean-field variables $\lim\limits_{t\rightarrow \infty}\rho_i(t)=\rho_i^*$ are obtained numerically once the system fully relaxes its diffusion component—summing eq. \eqref{eq:meanFieldDiffusion} over all three compartments. 
As anticipated earlier, it is in principle possible to extend the $SIR$ model we consider here to more complex contagion dynamics, but in this case, some of the new type of individuals (e.g., the ones that quarantine) might not contribute to the mobility, and this would unnecessarily complicate the derivation of eq. \eqref{eq:entropy}.  

Noe that we have a measure that globally quantifies the mobility of walkers in a given stochastic process, we turn our attention to the possibility that, while slowing down the spreading of infection, it is possible to keep a good amount of efficiency in the overall mobility. 
To address this question, we will explore the domain of parameters $\boldsymbol{\mathcal{D}}=\left(D, \boldsymbol{\sigma}\right)$ where $\boldsymbol{\sigma}=\left[\sigma_1, \sigma_2, \dots, \sigma_\Omega\right]$  and will simultaneously minimize the peak of infection $I_{max}$ and maximize the overall mobility $H$. Such multi-objective optimization,  known as Pareto optimization \cite{miettinen_nonlinear_1998}, can be mathematically formulated as follows:
\begin{eqnarray}
\min&& \left( I_{max}\left(D, \boldsymbol{\sigma}\right), H^{-1}\left(D, \boldsymbol{\sigma}\right) \right)\nonumber\\
&& s.t. \quad \left(D, \boldsymbol{\sigma}\right)\in \boldsymbol{\mathcal{D}}.
\end{eqnarray}
Notice that, in general, there is not a unique solution to the problem above. Instead, it can be shown that it exists a set of optimal outcomes denoted as the Pareto front \cite{miettinen_nonlinear_1998}. 
Our aim here is to find the points of the Pareto front in the NLRW case which when compared to the case where crowding is absent, i.e., $\sigma_i=0, \forall i$ (thus the diffusion is modeled as an LRW process), the peak of infection $I_{max}$ is lower for the same Entropy Rate $H$. 
From the mathematical point of view, we expect that by increasing the degree of freedom of the feasible set through the set of variables, $\sigma_1, \sigma_2, \dots, \sigma_\Omega$, we can access optimal points that are not possible otherwise. 
The drawback of incrementing the variable space lies in the high complexity of the Pareto optimization algorithm \footnote{In the following, we have made use of the \textsf{pareto()} function of the MATLAB Global Optimization toolbox.}, making it complicated to compute even for relatively small networks. 
To resolve this problem, we will constrain the search of the Pareto optimal points in a restricted subset of the parameters space. 
An immediate choice is to search for optimality in the parametric line $\sigma_i=\sigma$ where now $\sigma$ acts as the line parameter. Another possibility is to consider the parametric line $\sigma_i=k_i\sigma$ where in this case the line coefficient in the $\Omega-$dimensional space is equal the the node degree $k_i$. Thus in this later scenario, we consider more stringent measures for metanodes with a higher number of connections.

\begin{figure*}[h!]
    \centering
    \includegraphics[width=\linewidth]{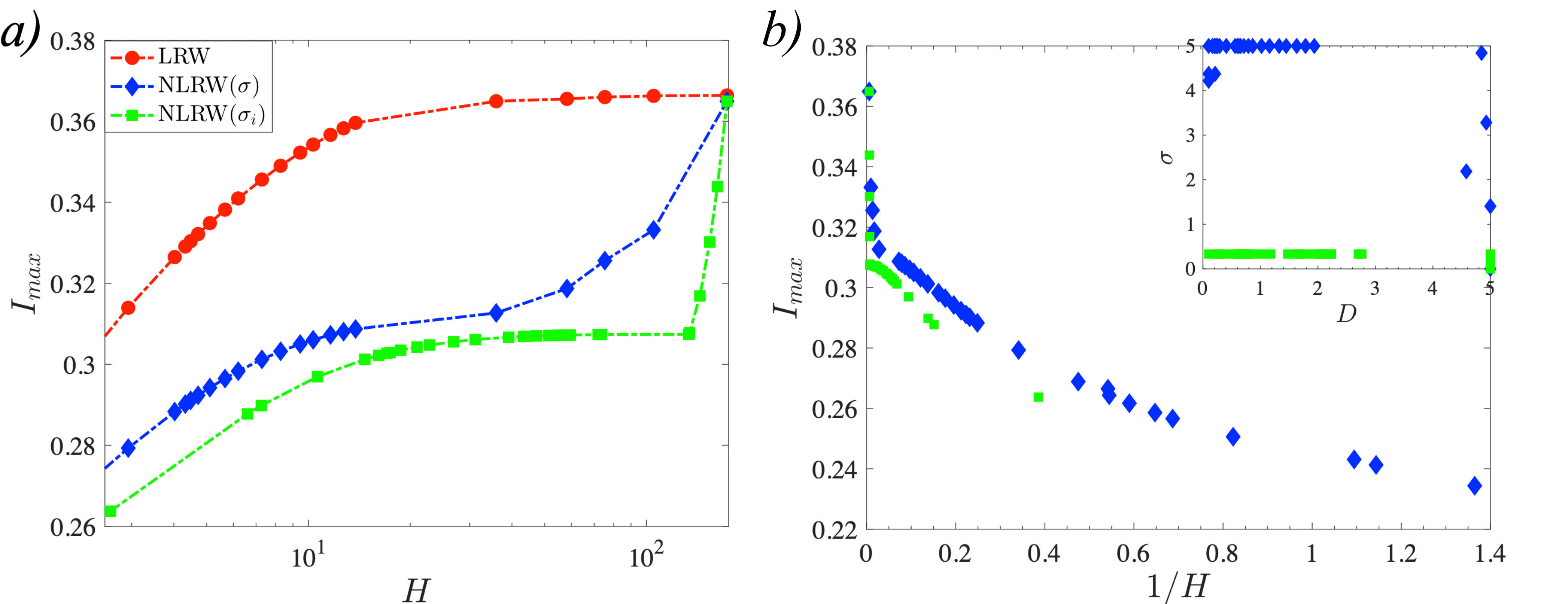}
    \caption{$\textbf{a)}$ Comparison of the evolution of minima of the infection peaks $I_{max}$ vs. the maxima of Entropy Rate (ER) $H$ between the spreading with Linear Random Walk (LRW) diffusion (red circles) and Nonlinear Random Walk (NLRW) diffusion (blue diamonds for a fixed $\sigma$ and green squares $\sigma_i$).  $\textbf{b)}$ For the NLRW case, the results were obtained calculating the Pareto front of the two-objective function $\left(I_{max}, 1/H\right)$ as a function of the Pareto set $(D, \sigma)$ (for both cases $\sigma_i=\sigma$ and $\sigma_i=\sigma\,k_i$) in the inset. Instead, the LRW curve was obtained by simply varying the diffusion rate $D$ by selecting the same ER values $H$. Thus there is only a single possible curve in the case of the LRW, as there is only a single parameter to change, D. For any value of $H$, the infection is always lower when the individuals avoid crowded nodes, and the simulations were carried with the same graph and parameters of Fig. \ref{fig:InfectionER}.}
    \label{fig:ER_crowd}
\end{figure*}

Based on such considerations, in Fig. \ref{fig:ER_crowd}, we show that the task of decreasing the infection while maximizing mobility is indeed possible. In fact, comparing the case of LRW diffusion (red curve) vs. both cases of the NLRW one (blue and green curves) for latent spatial network, one can immediately notice that the same values of the Entropy Rate $H$ have considerably lower maxima of infection. 
Furthermore such maximum values of infection are always lower when the crowding parameter $\sigma_i$ varies proportionally with the nodes' degree. 
This result shows that controlling the level of infection in a given population does not necessarily imply a restriction of mobility via partial or complete lockdown measures, but that better control on the capacity of all the spatially distributed sites where people can enter into contact with each other, is the optimal choice.

\begin{figure*}[h!]
    \includegraphics[width=\textwidth]{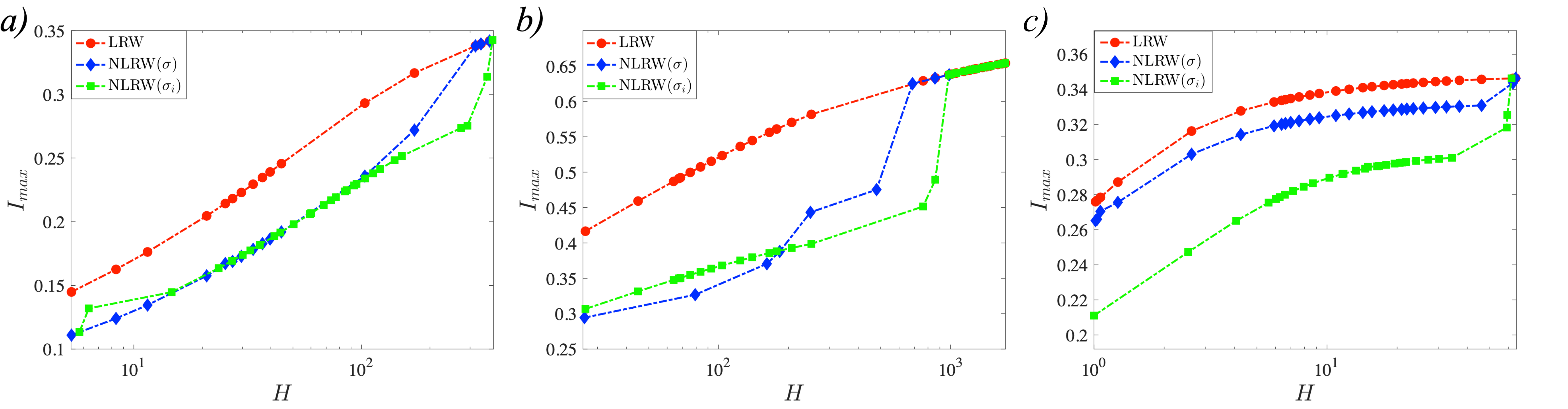}
    \caption{The minima of the infection peaks $I_{max}$ vs. the maxima of the Entropy Rate $H$ for the LRW (red circles) and NLRW (blue diamonds for a fixed $\sigma$ and green squares for a varying $\sigma_i$) mobility. The empirical spatial networks considered are $\textbf{a)}$ London Tube, $\textbf{b)}$ airlines route, and $\textbf{c)}$ USA contiguous map. In all cases the NLRW mobility performs as well as, and usually better than, the case of LRW diffusion. The parameters are $\lambda = 1$, $\gamma= 0.1$.}
    \label{fig:spatial_nets}
\end{figure*}

We show that the similar results can be obtained in a collection of empirical spatial networks --- the London Tube \cite{londonTransport}, airline routes \cite{pajek} and USA contiguous map network \cite{konect, USA} --- when we implement the Pareto optimization of the $SIR$ model with NLRW diffusion. 
In Fig. \ref{fig:spatial_nets}, it is shown that in each of these networks, the peak of infection decreases monotonically when the Entropy Rate $H$ decreases and the gap between the two approaches changes according to the different network topologies. 
For the airline routes network, for instance, we have the same peaks of infections for a very high level of mobility (no lockdown measures) in both LRW and NLRW cases, suggesting that a general reduction of mobility is needed for the proposed \textit{ad hoc} measures to be effective.\\ 
\onecolumngrid

\twocolumngrid 

\onecolumngrid

\twocolumngrid

\section{Conclusions}

In this study, we presented a reaction-diffusion model on a metapopulation network for optimal control of the infection spread, and tested it on synthetic networks and real-world data sets. 
This work aims to find a common approach to how human mobility should be regulated to keep the same pace of efficiency of human activity while keeping a low level of infection in the system. 
Starting from this premise, we have first reformulated the diffusion of individuals based on a (biased) Nonlinear Random Walk (NLRW) process introduced earlier in the literature \cite{Asllani2018, Carletti2020}. 
The latter differs from the classical Linear Random Walk (LRW) diffusion by considering a finite carrying occupancy per node, making the transition probability dependant on the density of individuals of the neighbour nodes. 
To further strengthen the social aspects of how individuals move through different spatial patches represented by interconnected vertices, we have introduced a node-based parameter that accounts for individuals' perception of the amount of crowding in a given node. 
Based on the master equation formalism, we show that a nonlinear deterministic transport operator is derived at the mean-field level. 
It is then possible to complement the formalism through a set of contagion interactions at the node level, where for definiteness, 
we have considered the paradigmatic Susceptible-Infected-Recovered model \cite{murray_mathematical_2002}.

Based on this formulation, we investigate the role that an NLRW diffusion has on the spreading of infection, and in particular, we show that not only the surge of disease is delayed, but also lower peaks can occur, and that the load of infection is distributed in several smaller waves. 
Indeed is possible to prove based on a linear prediction that crowding can slow down the spreading of the disease. 
However, at the first moment, this can intuitively be understood due to the reduction of general mobility. 
The latter is, of course, an undesired effect that will negatively impact not only the population from an economic and social aspect but also the access to the health system in general \cite{pike_economic_2014, fan_pandemic_2018}.
Nevertheless, the nonlinear diffusion model has a higher complexity regarding the number of parameters influencing the possible resulting scenario. 
In fact, starting from this perspective, we implement a multi-objective (Pareto) optimization on the control parameters, simultaneously aiming to determine the lowest possible peak of infection within the maximum possible mobility efficiency. 
The latter is formulated based on an information theory concept known as the Entropy Rate and quantifies the general level of performance of a stochastic process \cite{cover_elements_2006, Gomez-Gardenes2008}. 
We demonstrate that in the NLRW setting, it is possible to control the spreading of epidemics by choosing a set of parameters that reach the Pareto front, where the peak of infection is considerably lower than in the LRW case in which no restriction on the diffusion is considered. 
We have verified the validity of our result on both synthetic and empirical spatial networks. 

In conclusion, the mathematical model we develop in this paper can outline protocol measures for a targeted set of measures at the level of workplaces, schools, retail activities, hospitals, etc., by simply imposing a sufficient maximum occupancy on the latter. 
Our results suggest novel scenarios where mitigation policies can be based on a dynamically tuneable limitation of the capacity of indoor venues in order to contain and control the level of infection in the society and at the same time minimise the effect that such measures have on the human mobility. From this viewpoint, it is possible to better schedule the varying activities that require people to spend a long time in crowded and poorly ventilated spaces. This would allow a good level of human activity with less adverse effects of the ongoing epidemics.

\section*{Acknowledgements}
This work was partially supported by Irish Research Council [grant number GOIPG/2018/3026] (BAS) and by Science Foundation Ireland Grant numbers 16/IA/4470, 16/RC/3918 and 12/RC/2289P2 (JPG and MA).

\appendix
\section{Local analysis of the epidemic outbreak}
Here we use a linear stability analysis approach to predict the initial rate of infection growth for the NLRW diffusion operator. To proceed with the explicit calculations, we first need to assume the particular case of regular graphs, i.e., $k_i=k$, and $\sigma_i=\sigma, \forall i$.

Let us first recall that the NLRW Laplacian acting on the susceptible species for regular graphs is given by:
\begin{align}
   \sum_j L_{ij}(S)=D \sum_j \Delta_{ij}& \left[S_j\left(1-S_i-I_i-R_i\right)^\sigma\right. +\nonumber\\ &- \left.S_i\left(1-S_j-I_j-R_j\right)^\sigma \right].
\end{align}
Now consider perturbing from the initial steady state where no infected (and consequently no recovered) individuals are present, such that $S_i = S^* + \delta S_i$, $I_i = 0 + \delta I_i$, and $R_i = 0 + \delta R_i$, which gives
\begin{align*}
    D \sum_j \Delta_{ij} &\left[\left(S^* + \delta S_j\right)\left(1 - S^* - \delta S_i - \delta I_i -\delta R_i\right)^\sigma\right. + \nonumber\\
    &-\left. \left(S^* + \delta S_i\right)\left(1 - \delta S_j - \delta I_j - \delta R_j\right)^\sigma \right].
\end{align*}
Then we linearize the following term via a Taylor approximation,
\begin{align*}
    \left(S^* + \delta S_j\right)&\left(1 - S^* - \delta S_i - \delta I_i-\delta R_i\right)^\sigma \approx\\
    &\approx S^*\left(1-S^*\right)^\sigma + \delta S_j \left(1-S^*\right)^\sigma +\\
    &- S^* \sigma\left(1-S^*\right)^{\left(\sigma - 1\right)}\left(\delta S_i + \delta I_i + \delta R_i\right).
\end{align*}
Subbing this in gives
\begin{align*}
D \left[ {S^*\left(1-S^*\right)^\sigma \sum_j \Delta_{ij}}
    + \left(1-S^*\right)^\sigma\sum_j \Delta_{ij}\delta S_j + \right.\\
    - {S^*\sigma (1-S^*)^{(\sigma-1)}(\delta S_i + \delta I_i + \delta R_i)\sum_j \Delta_{ij}} +\\
    - {S^*(1-S^*)^\sigma \sum_j \Delta_{ij}} 
    - (1-S^*)^\sigma\delta S_i \sum_j \Delta_{ij}+\\
    + \left. S^* \sigma (1-S^*)^{(\sigma-1)}\sum_j \Delta_{ij}\left(\delta S_j + \delta I_j + \delta R_j\right)\right]
\end{align*}
All these terms, but the second one, vanish since the sum of the rows of the Laplacian $\Delta_{ij}$ is zero, and the sum of all the perturbations, in each node, equals zero \footnote{To be more rigorous due to the linear diffusion, the sum of the perturbation $\delta S_j + \delta I_j + \delta R_j$ does not vanish completely, but since the initial values of perturbation are small such contribution is negligible.}, which leaves us with
\begin{equation}
    D(1-S^*)^\sigma\sum_j \Delta_{ij}\delta S_j.
\end{equation}
The procedure above repeats similarly for all the species, thus fully justifying eq. \eqref{eq:lin_diff} of the main text.

\bibliographystyle{apsrev4-2}
\bibliography{mybib}

\end{document}